\documentclass[11pt,twocolumn,twoside]{IEEEtran}
\usepackage{amsmath}
\usepackage[margin=0.75in,headheight=0.45in]{geometry}
\usepackage[pdftex]{epsfig}
\usepackage{amsfonts}
\usepackage{amssymb}
\usepackage{fancyhdr}
\include{graphicsx}
\usepackage{bm}
 \graphicspath{{figures/}}
\begin{document}
\title{\vspace{0.2in} Identifying precipitation regimes in China using model-based clustering of spatial functional data}
\author{Haozhe Zhang$^{1}$, Zhengyuan Zhu$^{1}$\thanks{Corresponding author: Zhengyuan Zhu, zhuz@iastate.edu $^1$Department of Statistics, Iowa State University, Ames, IA $^2$State Key Laboratory of Earth Surface Processes and Resource Ecology $^3$School of Geography, Beijing Normal University, Beijing, China}, Shuiqing Yin$^{2,3}$}

\maketitle
\thispagestyle{fancy}
\begin{abstract}
The identification of precipitation regimes is important for many purposes such as agricultural planning, water resource management, and return period estimation. 
Since precipitation and other related meteorological data typically exhibit spatial dependency and different characteristics at different time scales, clustering such data presents unique challenges. In this short paper, we develop a flexible model-based approach to identify precipitation regimes in China by clustering spatial functional data. Though the focus of this study is on precipitation data, this methodology is generally applicable to other environmental data with similar structure.
\end{abstract}

\section{Introduction}
The study of precipitation in meteorology and climatology has a significant society impact. For example, drought and flood are two of the most serious meteorological disasters in China, with a direct economic loss of $177$ billion Chinese Yuan and annual average of 1256 deaths each year during the period 2001-2014~\cite{CMA2015}. Obvious seasonal and interannual variations of precipitation in China affected by Asian monsoon and complex terrain are the main reasons for the frequent drought and flood disasters. Dividing a large geographical area into more homogeneous precipitation regimes~\cite{li2009regionalization} has been shown to be useful for precipitation prediction, flood zone management, and regional extreme analysis~\cite{hosking2005regional}. Precipitation data has complex characteristics on multiple scales and typically has spatial and temporal dependence, which makes delineating precipitation regimes a non-trivial task. Motivated by this critical need, in this paper we develop a clustering approach for spatial functional data, and apply it to the precipitation data in China.

Regionalization problem has been studied extensively in the meteorological literature. The empirical orthogonal function (EOF) analysis has been widely used for regionalization problems in environmental science~\cite{white1991climate},~\cite{uvo2003analysis}, which is equivalent to principal component analysis in statistics. The EOF is used in \cite{li2009regionalization} to analyze the normalized monthly mean precipitation data from 1961 to 2006 at 400 stations and obtained a precipitation regionalization focusing on seasonal and interannual variations. However, the seasonal advance and retreat of the summer monsoon rain belt in East Asia behave in a manner with a step of 10-15 days~\cite{ding2004seasonal}, which can not be accurately described using monthly data, and daily rainfall data may be more useful to describe this summer monsoon effect accrurately. Due to the limitation of EOF method, unevenly distributed stations in space can significantly affect the loading patterns.  For example, the station density in the western and eastern parts of China is very different, therefore, some stations in the eastern part of China were ignored in the EOF analysis, which led to loss of information.  


The motivation of this research is to identify precipitation regimes in China using precipitation data. In this article, we propose a model-based approach to clustering spatial functional data by incorporating both spatial and geographic information in the procedure. In section \ref{Model}, we introduce the functional linear model for observed data and Markov model for cluster memberships with geographic covariates.  In section \ref{Results}, we apply the proposed method to precipitation data.

\section{Data}
\begin{figure}[ht]
	\centering
	\includegraphics[width=0.5\textwidth]{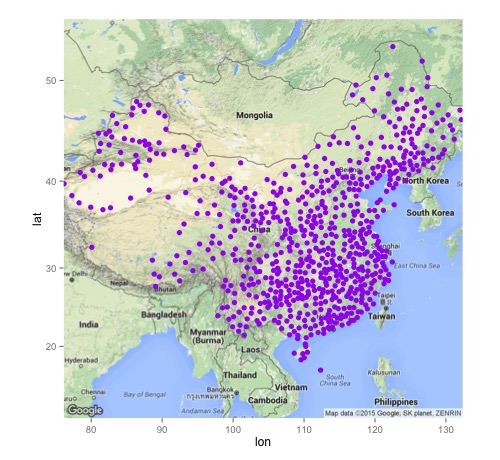}
	\caption{Spatial distribution of meteorological stations}\label{location}
\end{figure}

The data we analyze in this study is the daily precipitation data of 824 meteorological stations in the mainland China from 1951 through 2012. They were provided by the National Meteorological Information Center, China Meteorological Administration. The proportion of the missing days was $0.04\%$. Only those stations with more than 50 years' complete data are included in the analysis, so there are $722$ stations in total used in the analysis. The locations of these meteorological stations are shown in Fig \ref{location}.   

\section{Model}\label{Model}
Assume $Y_{ij}$ is the precipitation data observed in station $i$ at time point $t_{ij}$, where $i=1,\ldots, n$ and $j=1,\ldots, n_{i}$. Denote $\bm{Y}_{i}=(Y_{i1},\ldots,Y_{i,n_{i}})^{T}$. Let $Z_{i}$ be the cluster membership, called latent variable, following a multinomial distribution with support $\{1,\ldots\, C\}$. Here, $C$ is the number of clusters and is a tuning parameter. $Z_{i}=k$ if $\bm{Y}_{i}$ belongs to $k$th cluster. 
We call $\{(\bm{Y}_{i},Z_{i}): i=1,\ldots, n\}$ the complete dataset.
\subsection{Functional linear model for observed data}\label{flm}
Given the cluster membership $Z_{i}$, we assume $\bm{Y}_{i}|(Z_{i}=k)$ follows a multivariate normal distribution with a functional representation:
\begin{eqnarray}
\bm{Y}_{i}| (Z_{i}=k)=\bm{S}_{i}(\bm{\alpha}_{k}+\bm{\gamma}_{i})+\bm{\epsilon}_{i},\\
\bm{\gamma}_{i} \sim N(0,\bm{\Gamma}), \quad \bm{\epsilon}_{i} \sim N(\bm{0},\sigma^{2}I),
\end{eqnarray}
where $i=1,\ldots,n, k=1,\ldots,C$. In the functional linear model, $\bm{S}_{i}=(\bm{s}(t_{i1})^{T},\ldots ,\bm{s}(t_{in_{i}})^{T})^{T}$ is the basis matrix for $i$th curve. $\bm{s}(\cdot)$ is a vector of basis functions, which can be B-spline, Fourier or functional principal component. But the row number of basis matrix can vary across different curves to allow irregularly spaced time points and slight missing of data. $\bm{\alpha}_{k}$ is the coefficient and needs to be estimated. The data in the same cluster share the same coefficient $\bm{\alpha}_{k}$. The difference of $\{\bm{\alpha}_{k}\}$ reflects the heterogeneity across clusters. We assume the independence between distinct curves given cluster memberships. However, the within-curve dependence is accounted by the random effect $\bm{\gamma}_{i}$, since $\text{cov}(Y_{ij}, Y_{ij^{'}})=$the $(j,j^{'})$ element of $\bm{S}_{i}\bm{\Gamma}\bm{S}_{i}^{T}$. $\bm{\epsilon}_{i}$ can be regarded as the measurement error or stochastic error. Note that $\bm{\gamma}_{i}$ and $\bm{\epsilon}_{i}$ are confounded. Therefore, some constraint should be imposed for identifiability~\cite{james2003clustering}. We require that 
\begin{equation}
\bm{S}^{T}\bm{\Sigma}^{-1}\bm{S}=\bm{I},
\end{equation}
where $\bm{S}$ is the basis matrix evaluated over a fine lattice of time points that covers the full range of the data and $\bm{\Sigma}=\sigma^{2}\bm{I}+\bm{S}\bm{\Gamma}\bm{S}^{T}$.

\subsection{Markov model for cluster membership}\label{Markov}
To fully address the joint distribution of complete data $(\bm{Y}_{i},Z_{i})$, we need to specify the distribution of $Z_{i}$. Here, we assume the cluster membership follows a Markov model in space. We assume the following probability mass function of cluster memberships in the Markov model 
\begin{equation}\label{gibbs}
P(Z_{i}=k|\bm{Z}_{\partial i})=\frac{exp\{U_{ik}(\theta)\}}{N_{i}(\theta)}, 
\end{equation}
where $U_{ik}(\theta)=\theta \sum_{j \in \partial i}I(Z_{j}=k)$ is called the energy function and $N_{i}(\theta)=\sum_{k=1}^{C}exp\{U_{ik}(\theta)\}$ is the normalizing constant. $\theta$ is the interaction parameter that reflects the degree of interaction among nearby sites in Markov random field. The above distribution is called the Gibbs distribution~\cite{jiang2012clustering}, which originates from statistical physics but is widely used in spatial statistics.  

There are several ways to incorporate geographic covariates in the Markov model. One way is to generalize the definition of distance from Euclidean distance to "geographic distance" by spatial deformation. For instance, if there is a high mountain between two sites, then the distance between them can be set to be much larger than their euclidean distance on the earth but the geometric properties of Euclidean distance are still kept. The change of the definition of distance may lead to the respective change of neighbors. This method has been introduced in many papers in spatial statistics, to name a few, \cite{sampson1992nonparametric}, \cite{anderes2008estimating}, etc. The second way is to extend the energy distribution by imposing a function $f_{i,j}(\cdot)$ on $I(Z_{j}=k)$, i.e. $\tilde{U}_{ik}(\theta)=\theta \sum_{j \in \partial i}f_{i,j}\{I(Z_{j}=k)\}$ and $\tilde{N}_{i}(\theta)=\sum_{k=1}^{C}exp\{\tilde{U}_{ik}(\theta)\}$, where $f_{i,j}\{Z_{ik}\}$ is a function affected by the geographical covariates between site $i$ and one of its neighbors, i.e. site $j$. 

\section{Results}\label{Results}
\begin{figure}
	\centering
	\includegraphics[width=0.49\textwidth]{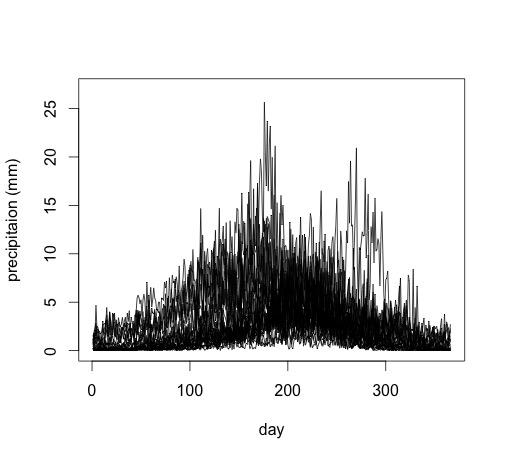}
	\caption{The averaged daily precipitation in some stations}\label{averaged daily precipitation}
\end{figure}

We applied this method to identify the precipitation regimes in China. Here, we focus on the interseasonal patterns of precipitation. The extension of this method to multi-scale functional and scalar data will be addressed in the following full paper. As a consequence, the averaged daily precipitation records within a year are used in the clustering. The detailed procedure of summarizing data is that, first we get the daily precipitation in each year from 1963 to 2012, then calculate the mean of these 50 curves. Some curves are illustrated in Fig \ref{averaged daily precipitation}. We used the second approach introduced in Section \ref{Markov} to incorporate geographical covariate. If the elevation difference between two stations is larger than 1000m \cite{gerlitz2015large}, we no longer consider them to be the ``neighbors`` in the Markov random field even if they are closest in terms of distance.

\begin{figure}
	\centering
	\includegraphics[width=0.49\textwidth]{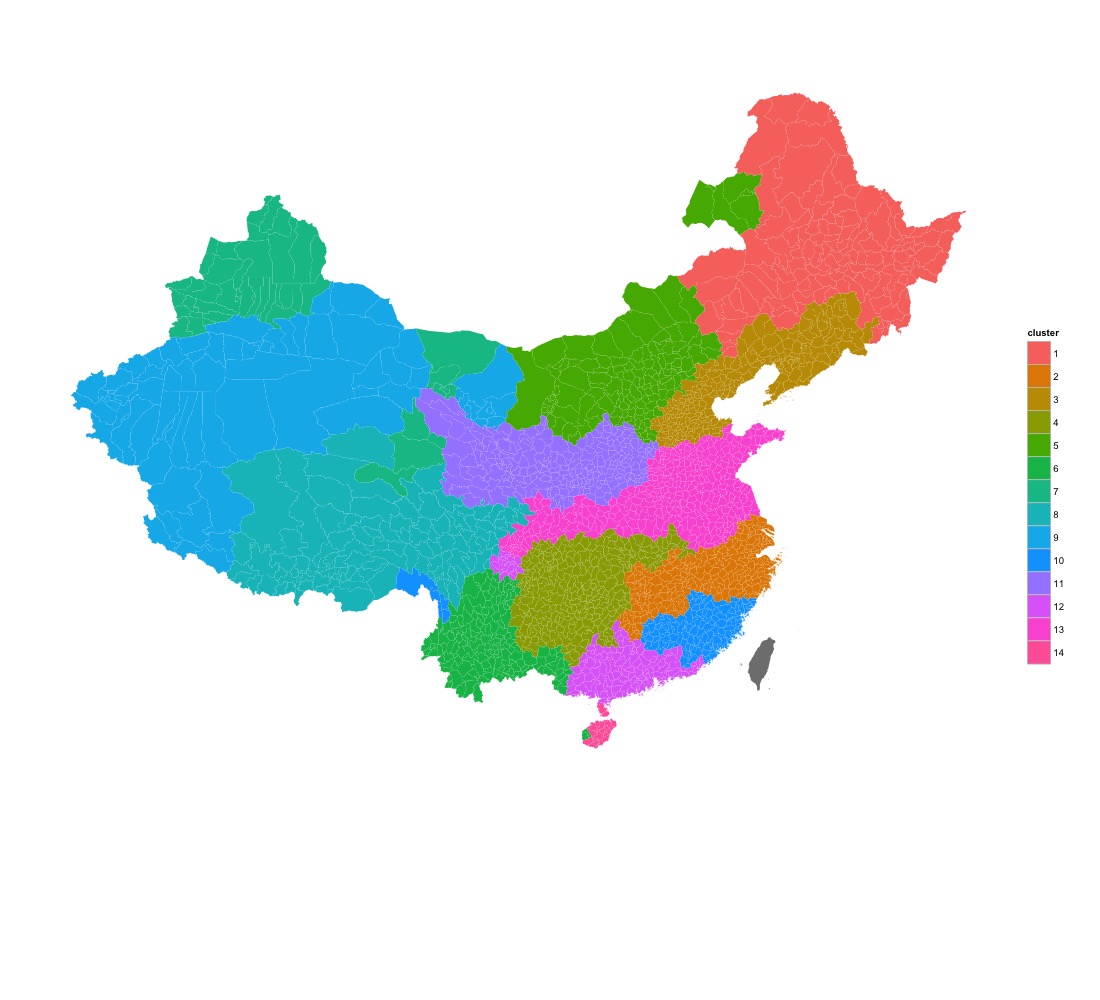}
	\caption{Regionalization of precipitation regimes in China}\label{clustering_map}
\end{figure}
The final cluster assignments are shown in Fig \ref{clustering_map}. The results of clustering are consistent with the stepwise manner of East Asian monsoon. The seasonal advance and retreat of the summer monsoonal airflow and monsoon rain belt in East Asia behave in a stepwise manner (Ding, 2004). When the East Asian summer monsoon advances northward, it undergoes three standing stages (South China and northern South China Sea from mid-May to early June; 25$-$30$^{\circ}$N from mid-June to mid-July; and 40$–-$45$^{\circ}$N during the last 10 days of July to mid-August), and two stages of abrupt northward shifts (the first 10 days of June and around mid-July). In early or mid-August the rainy season of North China comes to end, with the major monsoon rain belt disappearing. From the end of August to early September the monsoon rain belt moves back to South China again.

\section{Discussion}
In this short paper, we develop a flexible model-based approach to cluster precipitation data which utilizes the spatial and geographical information. There are still several important aspects of this method needed to be addressed, such as the selection of cluster numbers, how to evaluate the uncertainty of clustering assignments, etc. The parameter estimation, simulation study, model selection, extension to multi-scale data and uncertainty assessment will be introduced and addressed in the following full paper. 

\bibliographystyle{ieeetr}
\bibliography{ci_abstract}

\begin{thebibliography}{10}

\bibitem{CMA2015}
{\em China Meteorological Administration (CMA): Statistical Yearbooks of
  Meteorological Disasters in China}.
\newblock Meteorology Publishing House, 2015.

\bibitem{li2009regionalization}
C.~Li-Juan, C.~De-Liang, W.~Hui-Jun, and Y.~Jing-Hui, ``Regionalization of
  precipitation regimes in {C}hina,'' {\em Atmospheric and Oceanic Science
  Letters}, vol.~2, no.~5, pp.~301--307, 2009.

\bibitem{hosking2005regional}
J.~R.~M. Hosking and J.~R. Wallis, {\em Regional frequency analysis: an
  approach based on L-moments}.
\newblock Cambridge University Press, 2005.

\bibitem{white1991climate}
D.~White, M.~Richman, and B.~Yarnal, ``Climate regionalization and rotation of
  principal components,'' {\em International Journal of Climatology}, vol.~11,
  no.~1, pp.~1--25, 1991.

\bibitem{uvo2003analysis}
C.~B. Uvo, ``Analysis and regionalization of northern european winter
  precipitation based on its relationship with the north atlantic
  oscillation,'' {\em International Journal of Climatology}, vol.~23, no.~10,
  pp.~1185--1194, 2003.

\bibitem{ding2004seasonal}
Y.~Ding, ``Seasonal march of the east-asian summer monsoon,'' {\em East Asian
  Monsoon}, vol.~2, no.~30, p.~e53, 2004.

\bibitem{james2003clustering}
G.~M. James and C.~A. Sugar, ``Clustering for sparsely sampled functional
  data,'' {\em Journal of the American Statistical Association}, vol.~98,
  no.~462, pp.~397--408, 2003.

\bibitem{jiang2012clustering}
H.~Jiang and N.~Serban, ``Clustering random curves under spatial
  interdependence with application to service accessibility,'' {\em
  Technometrics}, vol.~54, no.~2, pp.~108--119, 2012.

\bibitem{sampson1992nonparametric}
P.~D. Sampson and P.~Guttorp, ``Nonparametric estimation of nonstationary
  spatial covariance structure,'' {\em Journal of the American Statistical
  Association}, vol.~87, no.~417, pp.~108--119, 1992.

\bibitem{anderes2008estimating}
E.~B. Anderes and M.~L. Stein, ``Estimating deformations of isotropic gaussian
  random fields on the plane,'' {\em The Annals of Statistics}, pp.~719--741,
  2008.

\bibitem{gerlitz2015large}
L.~Gerlitz, O.~Conrad, and J.~B{\"o}hner, ``Large-scale atmospheric forcing and
  topographic modification of precipitation rates over high asia-a
  neural-network-based approach,'' {\em Earth System Dynamics}, vol.~6, no.~1,
  p.~61, 2015.

\end{thebibliography}

\end{document}